    \documentclass[12pt,psf,epsf]{article}
\usepackage[centertags]{amsmath}
\usepackage{amssymb}
\usepackage{graphicx}
\usepackage{epsfig}
\usepackage{ulem}
\usepackage[english]{babel}
\usepackage{array}
\usepackage{amsthm}
\usepackage{latexsym}
\usepackage[mathcal]{euscript}
\pdfoutput=1
\usepackage{epsfig}

 \usepackage{jheppub}
 \usepackage{hyperref}
\usepackage{xcolor}

\usepackage{subfigure}
\usepackage{psfrag}

\usepackage{epsfig}
\usepackage[latin1]{inputenc}
\usepackage{float}
\usepackage{graphicx}
\usepackage{cancel}
\usepackage{mathrsfs}
\usepackage{amssymb}
\usepackage{amsfonts}
\usepackage{amsmath}
\usepackage{slashed}
\usepackage{graphicx}
\usepackage{bm}
\usepackage{color}
\newcommand{\be}{\begin{equation}}
\newcommand{\ee}{\end{equation}}
\newcommand{\bea}{\begin{eqnarray}}
\newcommand{\eea}{\end{eqnarray}}
\newcommand{\bwt}{\begin{widetext}}
\newcommand{\ewt}{\end{widetext}}
\newcommand{\cA}{A}

\newcommand{\nn}{\nonumber}

\newcommand{\bi}{\begin{itemize}}
\newcommand{\ei}{\end{itemize}}

\newcommand\cO{{\cal O}}

\newcommand\cM{{\cal M}}

\usepackage{setspace}

\begin{document}

\title {Sharp disentanglement in holographic charged local quench}

\author{Dmitry S. Ageev}
\affiliation{Steklov Mathematical Institute, Russian Academy of Sciences, Gubkin str. 8, 119991
Moscow, Russia}

\emailAdd{ageev@mi-ras.ru}

\abstract{We propose a charged falling particle in an AdS space as a holographic model of local charged quench generalizing model of arXiv:1302.5703. The quench is followed by inhomogeneous time-dependent distribution of chemical potential and charge. We derive the formulae describing the evolution of the entanglement entropy. At some characteristic time after the quench, we find a sharp dip in the entanglement entropy. This effect is universal and independent of the dimension of the system. At  finite temperature generalization of this model, we find that multiple dips and ramps appear during the evolution. }

\maketitle

\newpage

\section{Introduction}
The AdS/CFT  correspondence or holographic duality is an important and universal tool for the description of quantum phenomena.  It is a very useful theoretical tool  to  describe different phenomena in certain  quantum systems  by analyzing their classical gravitational duals. The striking example of such a phenomena is the entanglement in strongly coupled systems. The entanglement entropy is an important part of the holographic description of quantum systems \cite{Ryu:2006bv}-\cite{Swingle:2009bg}.  The examples of applications of the AdS/CFT correspondence include  physics of heavy-ion collisions \cite{CasalderreySolana:2011us,Arefeva:2014kyw}  and condensed matter theory \cite{Zaanen:2015oix,Hartnoll:2016apf}. An important feature of holographic duality is that the description of non-equilibrium quantum phenomena naturally emerges in terms of dynamical gravitational backgrounds (see \cite{Liu:2018crr} for review).

There are two exactly solvable canonical  examples of non-equilibrium  processes in the conformal field theory  that are testing grounds for many concepts in holography and physics of entanglement. These processes are called local and global quenches.   The global quench  is the process when the system follows unitary time evolution after a global perturbation. Examples of global quenches include  the instantaneous change of coupling constant, the homogeneous energy injection, and the boundary state quench (see \cite{Calabrese:2016xau} for review). From the holographic viewpoint, the global quench is described by the collapsing shell \cite{Danielsson:1999fa}-\cite{Balasubramanian:2010ce} and this black hole formation model can provide holographic setup for quite diverse type of physical situations including non-trivial initial states \cite{Ageev:2017wet}, hyperscaling \cite{Alishahiha:2014cwa,Fonda:2014ula} and chemical potential \cite{Caceres:2012em}\footnote{Also see another type of global quench with mildly broken translation invariance  \cite{ab1,ab2}.}.

The local quench is the process when the system evolves after the strongly localized perturbation. In this paper, our main focus will be on this type of quench. The canonical examples of solvable local quenches in 2d CFT which have the holographic description are joining/splitting geometric quenches and operator quench \cite{Calabrese:2007mtj}-\cite{DeJonckheere:2018pbi}. In the operator quench protocol, one perturbs the system by the insertion of the local operator at some point and at time moment $t_0$. The holographic description of this quench has been proposed in \cite{Nozaki:2013wia} and explored in different versions in \cite{Shimaji:2018czt}-\cite{DeJonckheere:2018pbi}. The model proposed in \cite{Nozaki:2013wia} consists of the point particle at the quench point $x=0$  falling into the bulk and deforming   the Poincare $AdS_3$. One can find the exact analytic expression for the metric satisfying the Einstein equations with this point-like source. Moreover, one can extend this description to the operator quench at the finite temperature \cite{Caputa:2014eta,Caputa:2015waa} and higher-dimensional quenches (at zero temperature)\cite{Nozaki:2013wia}. Note, that the exact dual of these higher-dimensional quench models on the CFT side is not known at the moment.

This paper is devoted to the extension of the model \cite{Nozaki:2013wia} to the case where particle perturbing the bulk carries $U(1)$ charge. This implies that the model is dual to the local quench by the charged operator. Point-like charged perturbation in the bulk creates the distribution of the Maxwell field in the bulk and in the boundary theory this corresponds to the inhomogeneous distribution of the chemical potential. The inhomogeneous chemical potential and charge oscillations in the holographic context have been studied in \cite{Blake:2014lva}. We calculate the chemical potential and charge dynamics following the quench corresponding to perturbed $AdS_3$ and higher dimensional extensions explicitly.  For the two-dimensional system the chemical potential first evolves as two localized lumps and after some time of evolution they  change their sign on the opposite.
An important result of \cite{Nozaki:2013wia} is the calculation of the entanglement entropy for some certain subregions. Moreover, a good approximation to the entanglement entropy that can be applied in the holographic non-equilibrium processes  has been proposed in \cite{Nozaki:2013wia}. As it was shown in \cite{Nozaki:2013wia} this approximation works very well as for lower-dimensional as for higher-dimensional quenches. We extend this calculation to our setup and find the universal effect in the evolution of the entanglement entropy. In zero charge case  for the interval  $x\in(-\ell,\ell)$ the entanglement entropy shows peaks around $t \approx \ell$. Turning on a non-zero charge of the operator leads to the sharp dip in the entanglement at this time. 
We find that this effect is also present in the higher-dimensional analog of this quench.

The finite temperature charged operator local quench is described by the charged particle falling on the BTZ black hole horizon \cite{Caputa:2014eta,Caputa:2015waa}. We  find that  turning on non-zero temperature changes the evolution picture. Sharp entanglement dip is preceded by a sharp peak. After this dip, the small smooth peak is also present.

These results are in line with other works devoted to a local charged quenches. This topic is still quite unexplored in contrast to zero charge  quenches.
In \cite{David:2017eno} the evolution of the higher-spin local quench has been studied in the context of  a particular higher spin theory. They observed  the effect similar to our dips in the entanglement, however, it is not clear whether there is a straightforward relation between them. 
The evolution of the quantum system after the injection of a  localized charge density   has been considered in \cite{Krikun:2019wyi}. In \cite{Krikun:2019wyi} the background where the local perturbation propagates is the static Reissner-Nordstrom black hole and the perturbation is considered in the probe limit. 

The paper is organized in the following way. In section \ref{sec:prel} we discuss general setup and preliminary information about the holographic description of local quenches.  The section \ref{sec:T0} is devoted to the calculation of the entanglement entropy  evolution following the local charged quench at $T=0$.
 In section \ref{sec:T} we discuss the case of $T>0$ local charged quench. The last section is devoted to concluding remarks.

\section{Preliminaries}\label{sec:prel}
\subsection{Holographic description of local quenches: zero charge}
In this section we remind the description of holographic neutral local quenches.
The metric of the form
\bea\label{BTZ0}
ds^2=-(1-\cM+R^2)d\tau^2+\frac{dR^2}{1-\cM+R^2}+R^2 d\phi^2,
\eea
 corresponds to  the solution of three-dimensional gravity with cosmological constant  known as the BTZ black hole for $\cM>1$. The values $\cM<1$ corresponds to the metric of the conical defect.  Both of these spaces on the CFT side can be described as the primary operator insertion (see, for example, \cite{Chen:2017yze}). According to the holographic dictionary this operator corresponds to the static massive particle in the center of global $AdS$. In 2d CFT this state can be expressed in the form 
 \be 
 |\psi\rangle=\cO(0)|0\rangle.
 \ee 
 To study the CFT on the line deformed by the primary operator one has to consider the Poincare patch deformed by the particle. Unlike in the global $AdS$ the particle cannot be completely static so the metric and the dual system become dynamical. The self-consistent equations of motion for the point particle interacting with the gravity are complicated and in general it is hard to solve them.

\begin{figure}[t!]
\centering
\includegraphics[width=5. cm]{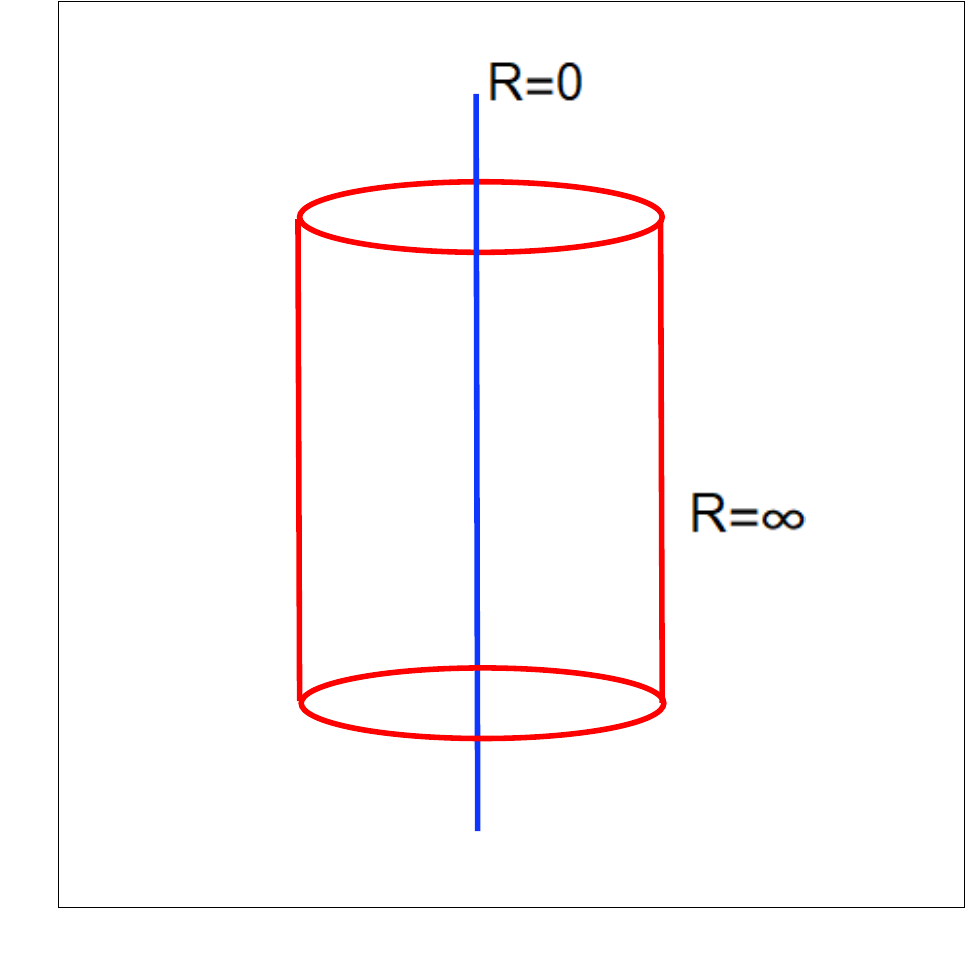}
\includegraphics[width=5.cm]{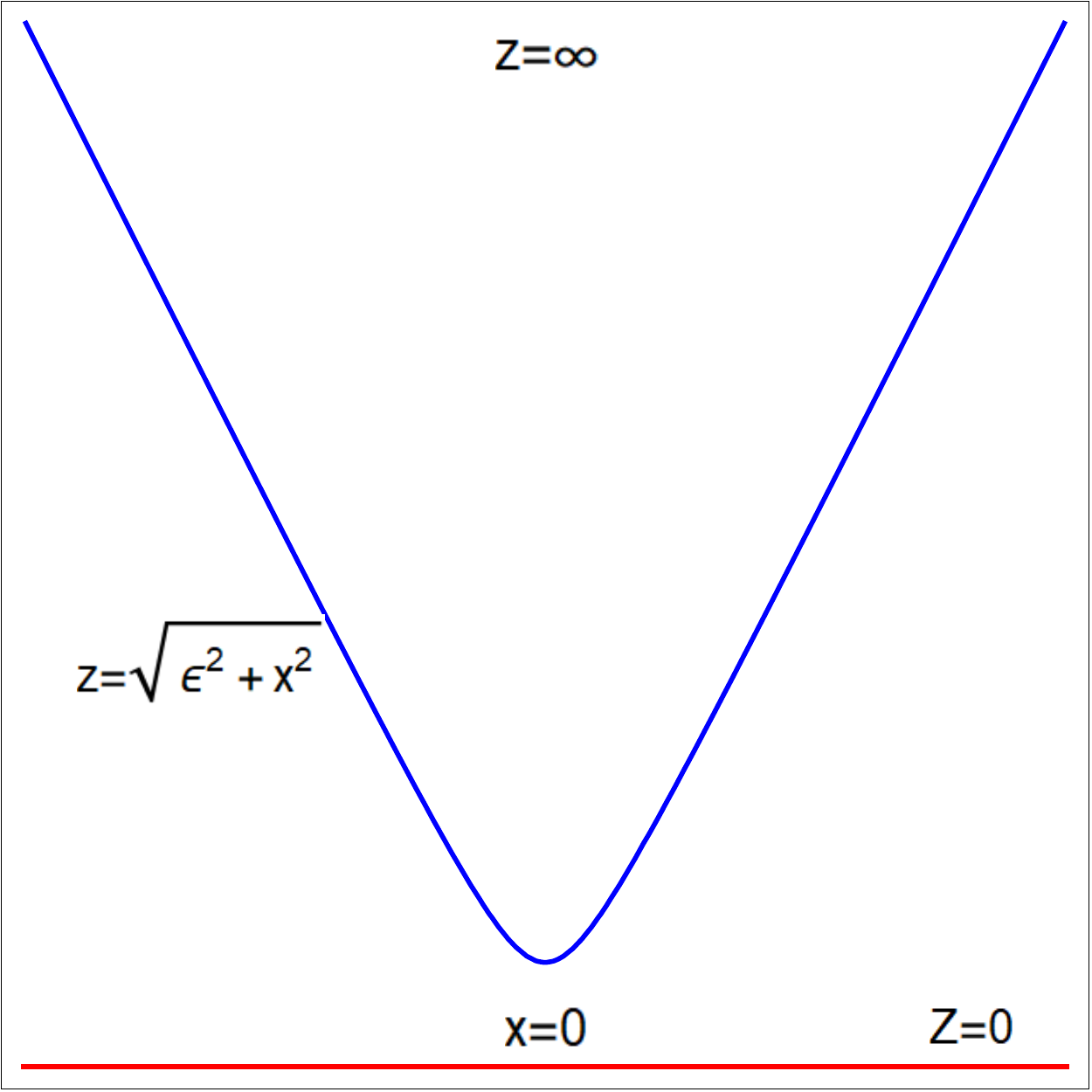}
\includegraphics[width=5.cm]{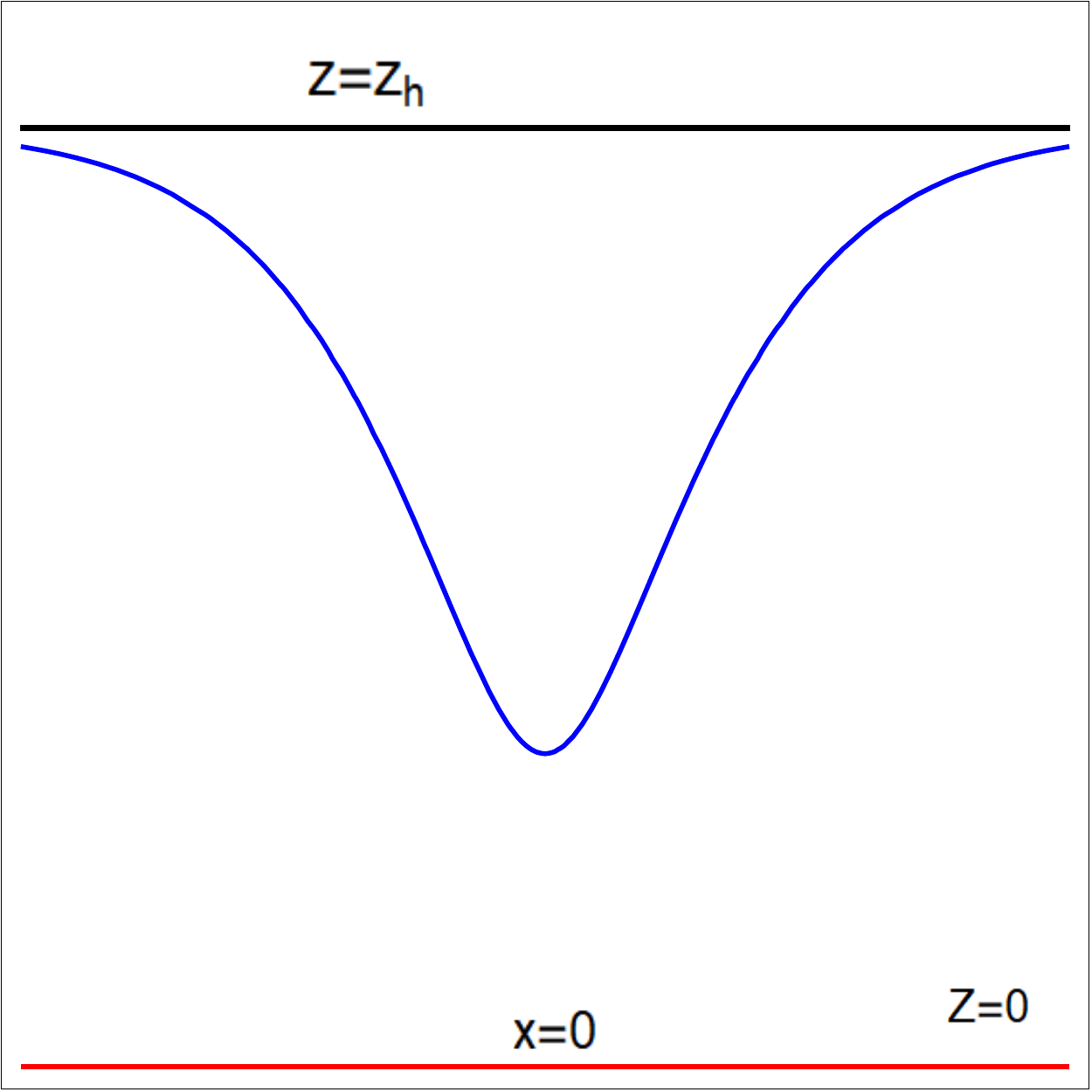}
 \caption{The schematic plot of particle and boundaries positions for global frame(right plot), poincare frame (central) and BTZ frame(right). Red curves correspond to the boundaries and blue one to the particle worldline (in different frames).}
 \label{fig:frames}
\end{figure}

However, we do not need to solve them in the straightforward manner. In \cite{Nozaki:2013wia} the following trick has been proposed. We know the metric for the static particle deforming the AdS in global coordinate frame. Then we make the change of variables to the Poincare patch. If $\cM=0$, this change of variables (see the explicit form below in the text, \eqref{mapzR}) transforms the metric \eqref{BTZ0} to the Poincare patch
\be 
ds^2=\frac{L^2}{z^2}\left(-dt^2+dx^2+dz^2\right).
\ee 
If $\cM>0$, the resulting metric has long and complicated form\footnote{For $Q=0$  one can find the explicit form of the metric in Appendix \ref{app:metric}.}, however it is still the metric with constant Ricci scalar except the particle worldline position.  The particle itself is mapped from $R=0$ in the initial metric  to the worldline $z=\sqrt{\alpha^2+t^2}$.   Also there is the generalization of this framework to the finite temperature case which we consider in Section \ref{sec:T}. The schematic plot of different frames with the perturbing particle is presented in Fig.\ref{fig:frames}.

In \cite{Nozaki:2013wia},  the generalization of the local  quench to higher-dimensional case also has been proposed. The static higher-dimensional version of \eqref{BTZ0} describing the geometry outside the massive object is given by
\bea \label{BH0}
&&ds^2=-f(R)dt^2+\frac{dR^2}{f(R)}+R^2d\Omega_{d-2}^2,\\
&&f(R)=1+R^2-\frac{\cM}{R^{d-2}},
\eea
which is the metric of higher-dimensional global $AdS_{d+1}$-Schwarzschild  black hole. After the change of variables in \eqref{BH0} from the global patch to the Poincare one we get the dual to $d$-dimensional CFT perturbed by insertion of the local operator. It was shown that the evolution of the system after such a perturbation leads to the radial waves emitted from the perturbation point and the qualitative picture is the same as in the lower dimensional theory \cite{Nozaki:2013wia}. 

In the case of three-dimensional gravity it is known that the BTZ black hole is locally AdS space. This implies that one can use this to construct the generalization of the operator local quench to finite temperature 2d CFT \cite{Caputa:2015waa}. Instead of mapping \eqref{BTZ0} to the Poincare frame one can map it to the BTZ frame. This leads to the picture where the particle is falling on the horizon and perturbing the static BTZ metric. 
$$$$
To introduce the chemical potential in the setting described above, one has to introduce the static geometry generalizing \eqref{BTZ0} and \eqref{BH0} first. The relevant metric has the form  
\bea\label{cBTZ0}
&&ds^2=-f(R)d\tau^2+\frac{dR^2}{f(R)}+R^2 d\phi^2,	\\
&&f(R)=1-\cM+R^2-\frac{1}{2}Q^2\log R,
\eea
with the gauge field given by
\be 
A_t=-(\Phi+Q\log R)dt,
\ee 
where $\Phi$ is the integration constant.
Typically $\Phi$ is set to be $\Phi=-Q\log r_h$ and $1-\cM=-r_h^2$. This metric is known as charged BTZ black hole with the horizon at $R=r_h$  and  $A_t$ is vanishing at $r_h$.  In higher-dimensional case the  geometry of interest is the AdS-Reissner-Nordstrom black hole defined by the function $f(R)$ of the form
\be \label{blc}
f(R)=1+R^2-\frac{M}{R^{d-2}}+\frac{Q^2}{R^{2(d-2)}}.
\ee 
and with the element $d\Omega_{d-1}^2$ instead $d\phi^2$.

In Appendix \ref{sec:statApp} we consider the behaviour of the geodesics in the metrics defined by \eqref{cBTZ0} and \eqref{blc}. The length of such geodesics (spanned on the boundary interval of the length $\ell$) corresponds to the entanglement entropy  for $d=2$ and for correlation function for $d>2$. We find that even for values of $Q^2<0$ this length is real-valued. The range of applicability and meaning of values  $Q^2<0$ is not completely clear. We discuss this issue in Appendix \ref{sec:statApp}. For completeness we include these values of $Q$ in discussion of nonequilibrium entanglement entropy in the next sections, however our main discussion will be about $Q^2>0$ corresponding to real-valued chemical potential.
\section{Quench by the charged operator at zero temperature}\label{sec:T0}
\subsection{2d CFT: charged BTZ black hole}
As it was mentioned in the previous section the metric corresponding to the $AdS_3$ perturbed by the static point charge has the form
\bea\label{cBTZ}
ds^2=-(1-\cM+R^2-\frac{1}{2}Q^2\log R)d\tau^2+\frac{dR^2}{1-\cM+R^2-\frac{1}{2}Q^2\log R}+R^2 d\phi^2,
\eea
 with the gauge field  
\be \label{btzgauge}
A_t=-Q\log \frac{R}{R_h} \,dt,
\ee 
where $R_h$ is some constant.
Global coordinates and the Poincare patch are related by the following coordinate transformation
\bea\label{mapzR}
&&z=\frac{\alpha }{R \cos (\phi )+\sqrt{R^2+1} \cos (\tau )},\\
&&t=\frac{\alpha  \sqrt{R^2+1} \sin (\tau )}{R \cos (\phi )+\sqrt{R^2+1} \cos (\tau )},\\
&&x=\frac{\alpha  R \sin (\phi )}{R \cos (\phi )+\sqrt{R^2+1} \cos (\tau )},
\eea
with the inverse mapping given by
\bea\label{mapRz}
&&\tau =\text{arctan}\left(\frac{2 \alpha  t}{\alpha ^2-t^2+x^2+z^2}\right),\\
&&R=\frac{\sqrt{\alpha ^4+2 \alpha ^2 \left(t^2+x^2-z^2\right)+\left(-t^2+x^2+z^2\right)^2}}{2 \alpha  z},\\
&&\phi =\text{arctan}\left(\frac{2 \alpha  x}{\alpha ^2+t^2-x^2-z^2}\right).
\eea
This mapping relates the worldline of the static particle $R=0$ and the worldline $z=\sqrt{\alpha^2+t^2}$ of the particle falling into the bulk of Poincare $AdS_3$ bulk and deforming it.
The gauge field corresponding to this dynamical background can be obtained in the straightforward manner also applying the mapping \eqref{mapRz} to the static gauge field \eqref{btzgauge}. The explicit form of the gauge after the mapping can be found in the Appendix \ref{sec:gauge}. 

\begin{figure}[t!]
\centering
\includegraphics[width=8cm]{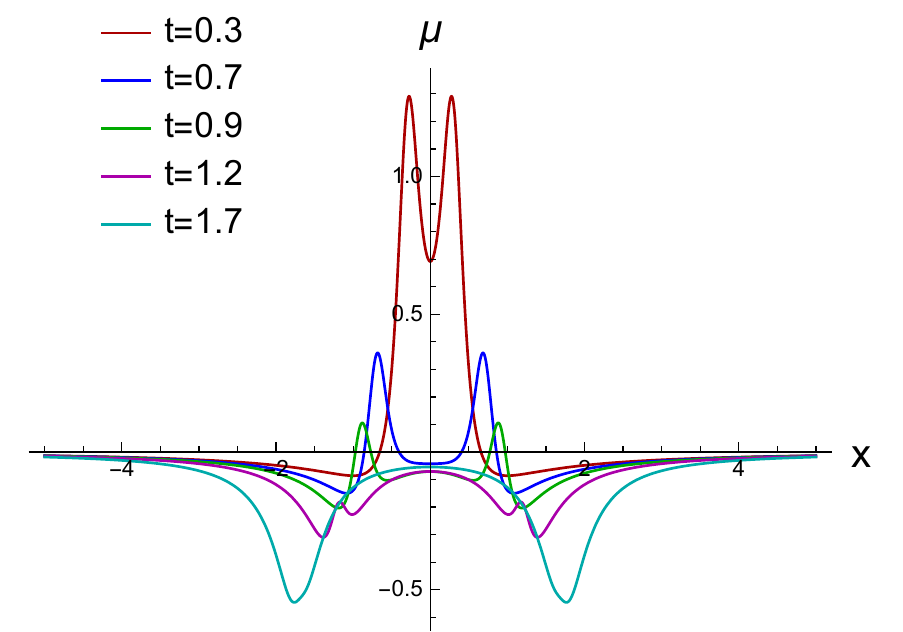}
\includegraphics[width=6.5cm]{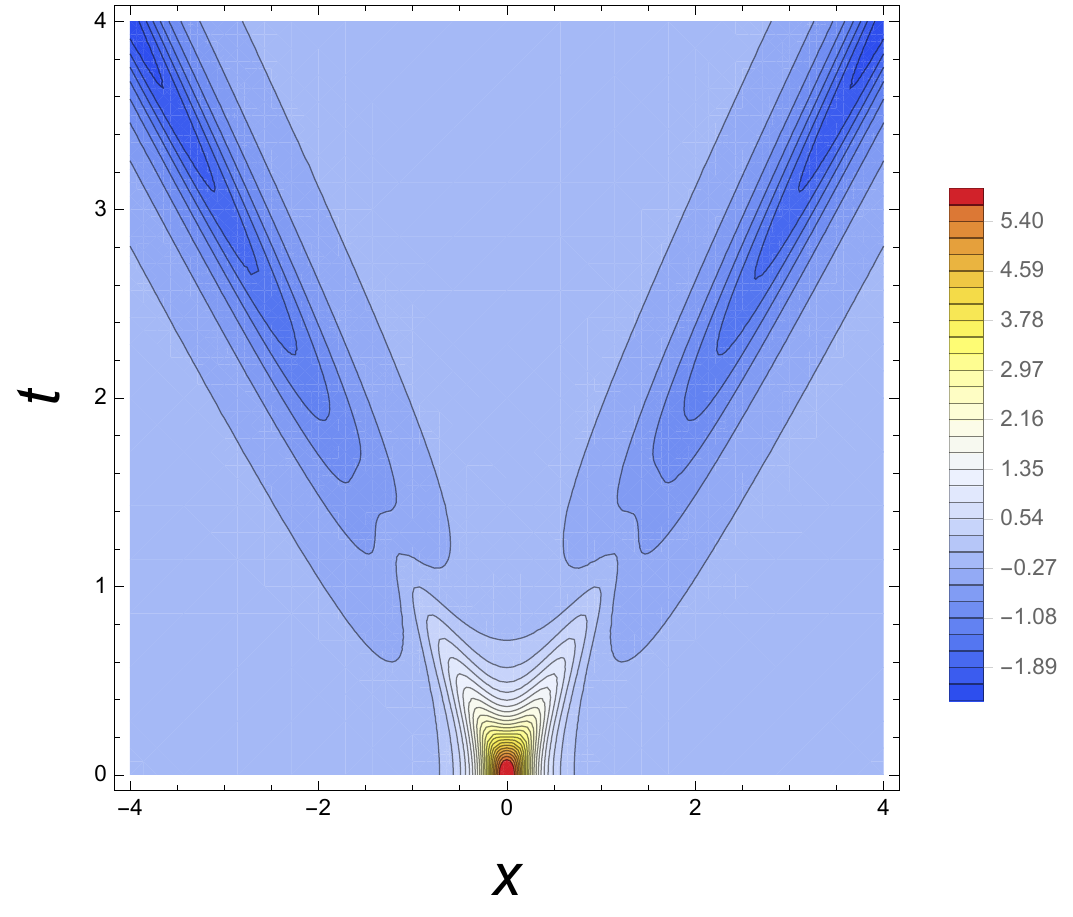}
 \caption{Left plot: chemical potential potential distribution given by \eqref{mu0} after the charged local quench with $\alpha=0.2$ and $Q=-0.3$ at different time moments. Right plot: the contour plot for the same quantity.}
 \label{fig:MU0}
\end{figure}

The solution with the metric \eqref{cBTZ} and corresponding gauge fields have the logarithmic terms which are absent in $d>2$ case. In \cite{Jensen:2010em}, it was shown that in the Einstein-Maxwell theory on $AdS_3$ for the near-boundary expansion of the gauge field of the form
\be 
A_\mu\sim a_\mu(x) \log r + b_\mu(x)+...,
\ee 
the components of the functions $a_t(x), b_t(x)$ are identified with the dynamical charge $q$ density and the chemical potential $\mu$ as 
\bea \label{prescr}
&&q=q(x,t) = a_t(x,t),\\
&&\mu=\mu(x,t)=b_t(x,t),
\eea 
and the current is identified with the function
\be \label{prescr2}
j_x(x,t) = a_x(x,t).
\ee
Note, that $A_z$ component of the field \eqref{dyngauge} vanishes near the boundary as it should be. Expanding the gauge field from Appendix \ref{sec:gauge} and taking \eqref{prescr}  we get the current  after the perturbation in the form
\bea 
 j_x(x,t)= -\frac{4 \alpha  Q t x}{\left(\alpha ^2+t^2\right)^2+2
   x^2 (\alpha^2-t^2)+x^4},
\eea 
and the dynamical chemical potential distribution
\be\label{mu0} 
\mu(t,x)=-\frac{2 \alpha  Q \left(\alpha ^2+t^2+x^2\right) \log
   \left((2 \alpha R_h)^{-1} \sqrt{\left(\alpha
   ^2+t^2\right)^2+2 x^2 (\alpha^2 -t^2)+x^4}\right)}{\left(\alpha ^2+t^2\right)^2+2 x^2
   (\alpha^2-t^2)+x^4}.
\ee 
We present the evolution of chemical potential given by \eqref{mu0} in Fig.\ref{fig:MU0}. One can observe formation of two localized perturbations which decrease during the evolution. After some time both perturbations change their sign and continue to propagate  with the sing opposite to the initial one. The presence of the dynamical chemical potential is unusual, but studied previously in the holographic context, for example, in \cite{Blake:2014lva,Krikun:2019wyi}.

$\,$

Applying the change of variables \eqref{mapRz}  to \eqref{cBTZ} we obtain the  metric dual to the charged local quench, which is the solution of the Einstein equations with the charged matter dual.   For convenience, we are not going to present this metric here (for $Q=0$  one can find the explicit form of the metric in Appendix \ref{app:metric}) because of the cumbersome form.
Now  turn to the computation of the entanglement entropy evolution following the charged local quench. In this case, the Hubeny-Ryu-Rangamani-Takaynagi (HRRT) prescription 
\be 
S=\frac{A}{4G}
\ee 
states that the entanglement entropy of the subsystem is given by the area of an extremal codimension-two hypersurface $A$ area (geodesic for three-dimensional gravity).
It was shown in \cite{Nozaki:2013wia} how to compute qualitatively (and for small expansion parameter $M$ even quantitatively) good approximation to the length of the geodesic connecting two equal-time points on the boundary. This approximation is constructed as follows.
Assume, that we have the expansion of the metric in the form 
\be 
g_{\mu\nu}\approx g_{\mu\nu}^{(0)}+g_{\mu\nu}^{(1)}+O(parameters)+...,
\ee 
where  ``parameters'' in our case are $M$ or $Q$. For the unperturbed metric  $g^{0}_{\mu\nu}$ the HRRT surface has the form
\be 
X_{\mu}^{(0)}=X_{\mu}(\xi_\nu),
\ee 
and the induced metrics are  defined as
\be \label{eeg}
G_{\alpha\beta}^{(0)}=\frac{\partial X^{\mu}}{\partial \xi^{\alpha}}\frac{\partial X^{\mu}}{\partial \xi^{\beta}} g_{\mu\nu}^{(0)},\, \,\,\,\,\,\,\,\, G_{\alpha\beta}^{(1)}=\frac{\partial X^{\mu}}{\partial \xi^{\alpha}}\frac{\partial X^{\mu}}{\partial \xi^{\beta}} g_{\mu\nu}^{(1)}.
\ee
Finally, the leading order perturbation  of the entanglement entropy over the state with $Q=0$ (normalized for convenience on the factor $1/(4G)$) has the form
\be \label{apprS}
\Delta S=\frac{1}{2}\int d^{d-1}\xi \sqrt{G^{(0)}}\text{Tr}\Big[G^{(1)} (G^{(0)})^{-1}\Big].
\ee 
For small $M$ and $Q$, this method gives good approximation (see also \cite{Ageev-quench,Ageev-quench-2} for applications of this method in the context of holographic complexity). This method is also applicable in the setting of the higher-dimensional local quenches.

\begin{figure}[t!]
\centering
\includegraphics[width=9.5cm]{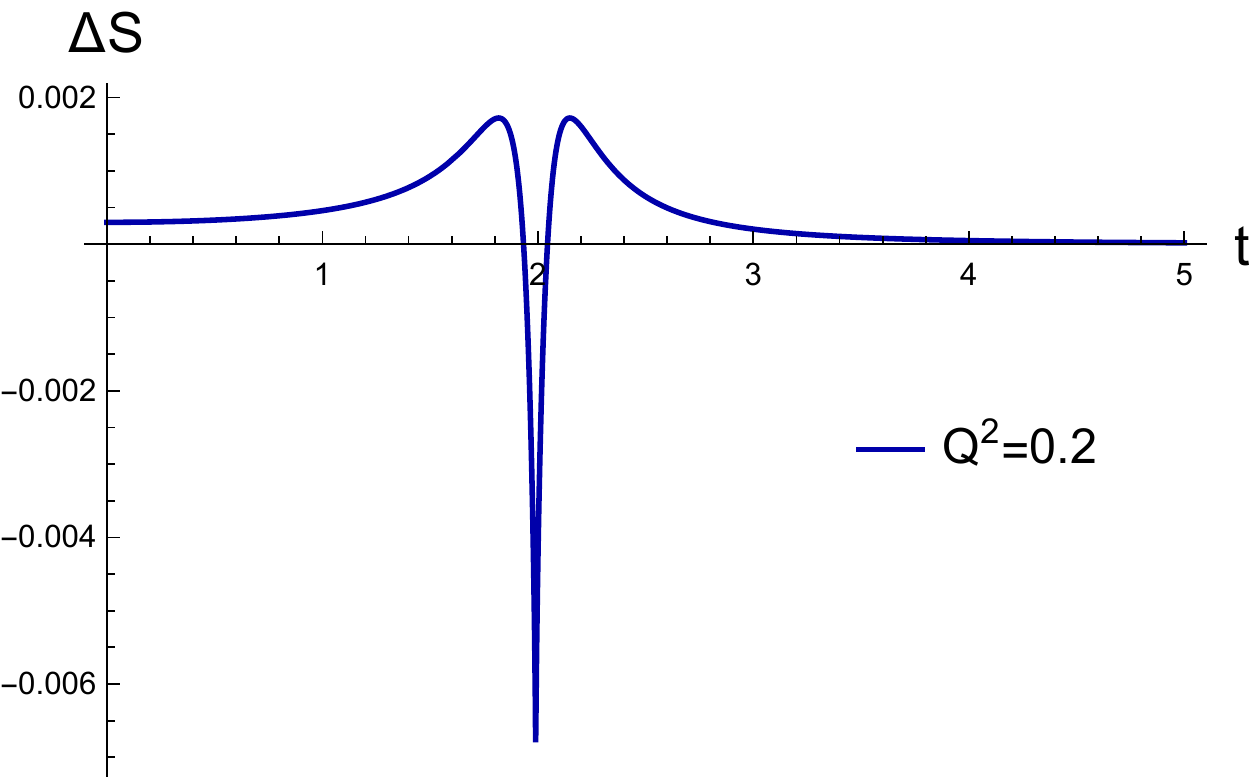}
 \caption{The evolution of $\Delta S$ given by \eqref{eeg2} for $Q^2=0.2$ and $\alpha=0.3$, the subsystem $x\in (-2,2)$ and $\cM=0.3$. }
 \label{fig:EEQ1}
\end{figure}

Applying this method, it is straightforward to calculate the entanglement entropy for the single interval after the quench. Using \eqref{eeg} and \eqref{apprS} with \eqref{cBTZ}  we get the integral expression for the evolution of the entanglement entropy perturbation
\be \label{eeg2}
\Delta S =- Q^2\int_{-\ell/2}^{\ell/2} \frac{\alpha^2  x^2 }{ 
   \left(\left(\alpha ^2+t^2-\ell ^2\right)^2+4 \alpha
   ^2 x^2\right)}\log \left(\frac{4 \alpha ^2
   \left(\ell ^2-x^2\right)}{\left(\alpha ^2+t^2-\ell
   ^2\right)^2+4 \alpha ^2 x^2}\right) dx,
\ee 
for the interval $x\in(-\ell/2,\ell/2)$.
Performing the integration numerically we obtain  the following picture of the entanglement evolution. The summary of the operator charge  effect  is the following
\begin{itemize}
    \item The entanglement evolution picture after the chargeless local quench is well known and we briefly remind it here. For the interval centered around the quench point the entanglement starts to increase, attaining the  sharp peak around  $t\approx \ell/2$, which is the time when quasiparticles reach the boundary of the interval.
    \item In Fig.\ref{fig:EEQ1} we present the evolution of the entanglement perturbation  $\Delta S$ for the charged quench with the fixed  $Q$. One can clearly see that for $Q^2>0$  there is the sharp dip at the same time where chargeless quench has peak.  At the early times $\Delta S$  is a slowly growing quantity.
    \item If we admit  $Q^2<0$ corresponding to the imaginary chemical potential  we have the opposite behaviour which is easily seen from \eqref{eeg2}. Thus instead of sharp dip we get the sharp peak and initially the entanglement is decreasing.
\end{itemize}
To understand the qualitative picture of the process, it is useful to look at the bulk picture of perturbation caused by the particle. We present the determinant of the metric on the constant time slice in Appendix \ref{sec:gauge}. In Fig.\ref{fig:dens} we see that after some time particle position is surrounded by the negative ``metric perturbation''  and there is also positive contribution of another localized perturbation propagating closer to the boundary. So the dynamics of the entanglement can be explained from this competition between these contributions to the ``metric density''.

\begin{figure}[t!]
\centering
\includegraphics[width=10.5cm]{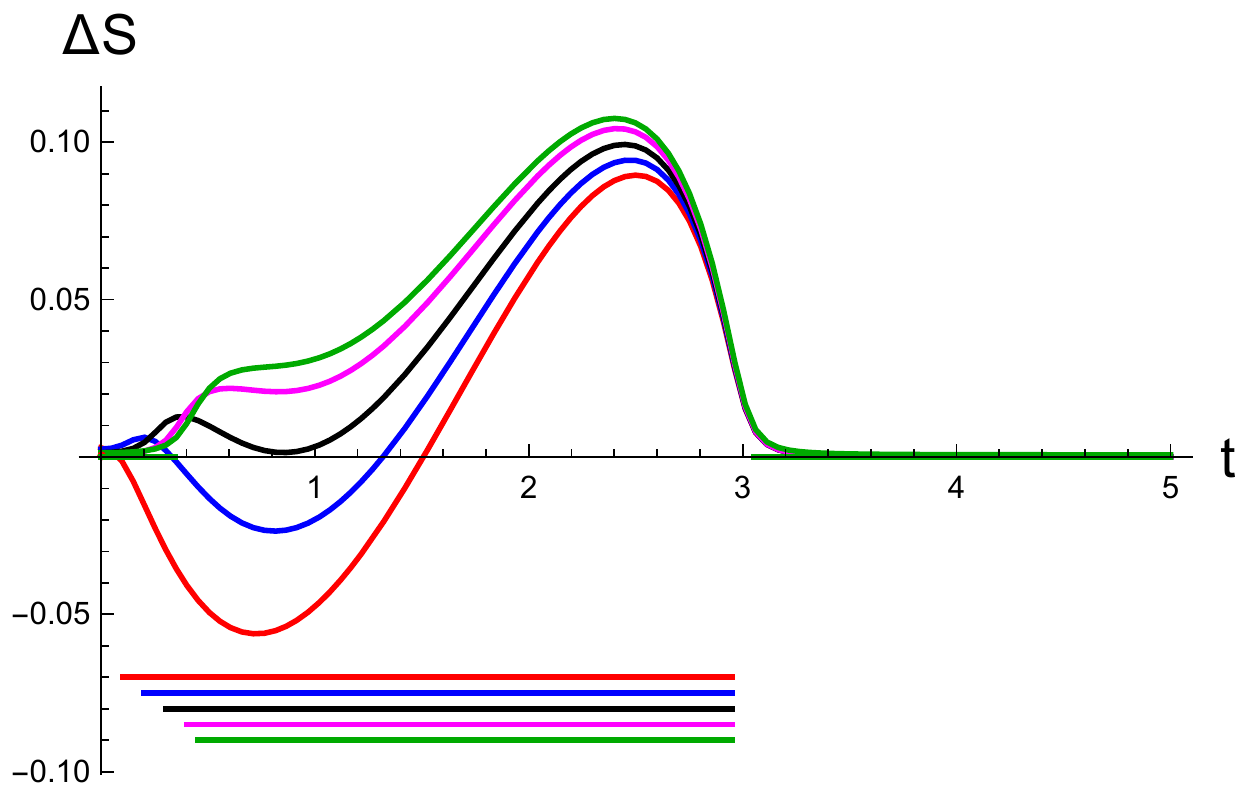}
 \caption{The evolution of entanglement perturbation $\Delta S$ for the interval $x\in(x_1,x_2)$ where $x_2$ is fixed.  Curves of different color are in one to one correspondence with the coloring of the corresponding intervals in the bottom of the plot. Here $Q=0.2$, $\cM=0.3$ and $x_2=3$.}
 \label{fig:EEQshft}
\end{figure}

This picture leads to quite sensitive behaviour of the entanglement entropy for some cases. The entanglement entropy evolution for the shifted interval $x\in(x_1,x_2)$ and $x_2>x_1>0$ is presented in Fig.\ref{fig:EEQshft}. One can see how small change of the left endpoint of the interval leads to the significant change in the behaviour of the entanglement.  The size of the dip depends on how close is the left endpoint to the point of the quench.
\subsection{Local charged quench at higher dimensions: ``falling'' Reissner-Nordstrom}
Now turn to the higher-dimensional generalization of local quench given in the previous section. Zero charge analog of higher-dimensional local quench has been developed in \cite{Nozaki:2013wia}. This construction has been proposed first in \cite{Horowitz:1999gf} and is called sometimes ``falling black hole''. In this setup, one  applies the construction from the previous subsection to the  global AdS-Schwarzschild black hole. For the charged generalization of higher-dimensional local quench one replaces this black hole with the global AdS-Reissner-Nordstrom black hole.
The global charged black hole solution in $d+1$ dimensions is given in the form
\bea
&&ds^2=-f(R)dt^2+\frac{dR^2}{f(R)}+R^2d\Omega_{d-1}^2,\\
&&f(R)=1+R^2-\frac{M}{R^{d-2}}+\frac{Q^2}{R^{2(d-1)}},\\
&&A_t=\left(\sqrt{\frac{d-2}{2(d-2)}} \frac{Q}{R^{d-3}}+\Phi \right)dt,
\eea 
where $\Phi$ is the constant typically chosen such, that the gauge field vanishes on the horizon. Let us focus  for definiteness on $d=4$ case (i.e. dual to the local perturbation on the plane).
The transformation to the Poincare patch in a general dimensions has the form
\bea 
&&R=\frac{\sqrt{\alpha ^4+2 \alpha ^2 \left(\rho
   ^2+t^2-z^2\right)+\left(\rho ^2-t^2+z^2\right)^2}}{2 \alpha  z},\\
&&\nn\tau=\text{arctan}\left(\frac{2 \alpha  t}{\alpha ^2+\rho
   ^2-t^2+z^2}\right),\\\nn
&&\phi=\text{arctan}\left(\frac{\alpha  x}{\sqrt{\alpha ^2
   \left(x^2+y^2\right)}},\frac{\alpha  y}{\sqrt{\alpha ^2
   \left(x^2+y^2\right)}}\right),\\\nn
&&\theta=\text{arctan}\left(\frac{-\alpha ^2+\rho ^2-t^2+z^2}{2 \alpha  \rho }\right).
\eea 
After the transformation to the Poincare coordinates and taking the near-boundary expansion of the gauge-field components (let us consider $d=3$, for definiteness) we obtain the current and the dynamical chemical potential after the perturbation
\bea
&&\mu(t,x)=\frac{2 \alpha  \Phi \left(\alpha ^2+\rho ^2+t^2\right)}{\left(\alpha
   ^2+\rho ^2\right)^2+2 t^2 (\alpha^2 -\rho^2 ) +t^4},\\
&&J_{\rho}(t,\rho)=-\frac{4 \alpha  \Phi \rho  t}{\left(\alpha ^2+\rho ^2\right)^2+2 t^2
   (\alpha^2 -\rho^2 )+t^4},
\eea 
where we introduced the radial coordinate  $\rho=\sum_i^d\sqrt{x_i^2}$ (in case of $d=2$ we have  $\rho=\sqrt{x^2+y^2}$). As in the case of one spatial dimension quench, we have the time-dependendent current and chemical potential with the spherical symmetry. For arbitrary $d$, the answer is essentially the same up to some prefactor.

$\,$

\begin{figure}[t!]
\centering
\includegraphics[width=10.5cm]{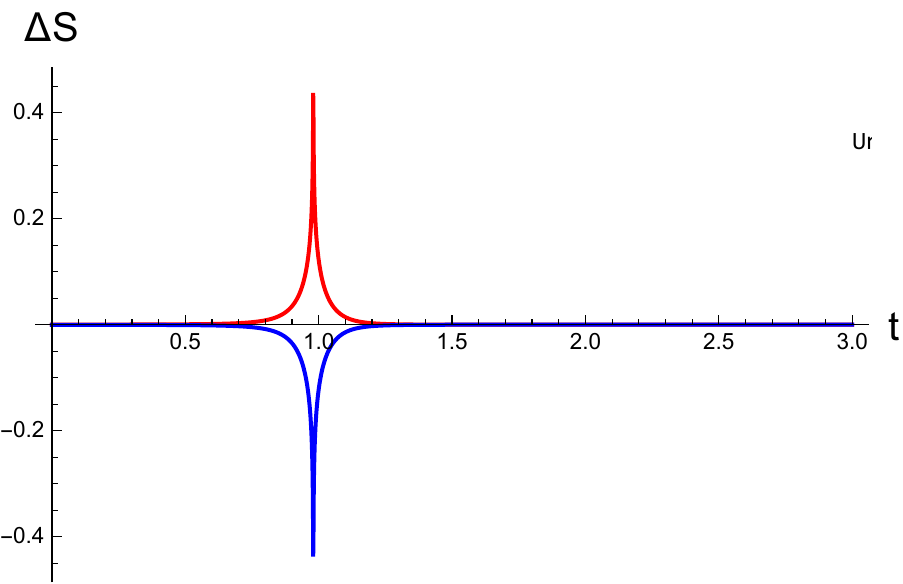}
 \caption{The entanglement entropy $\Delta S$ evolution, for $d=4$ for $Q=0.4$ entanglement and the other parameters being fixed. The evolution is given for the interval $(-2,2)$, $M=0.4$, $\alpha=0.25$.}
 \label{fig:EEQ3d-quench}
\end{figure}

Finding the exact HRRT surface in many dimensional case is complicated numerical problem  and to get the insight in the entanglement structure we use the same method as in the previous section. We choose the subsystem to be the region bounded by the circle of radius $R$.  The HRRT surface  in this case has the form
\be 
z(\rho)=\sqrt{\ell^2-\rho^2}.
\ee 
Repeating all calculation steps of the entanglement entropy as in $d=2$ case and using   \eqref{eeg2} we get the integral expression for the entanglement entropy perturbation in the form 
\be 
\Delta S \sim 2 \int_{0}^{\ell}\frac{8 \alpha ^4 Q^2 \rho ^3 \sqrt{(\ell^2 -\rho^2 )}}{\ell  \left(\alpha ^4+2 \alpha ^2 \left(2 \rho ^2+t^2-\ell
   ^2\right)+\left(\ell ^2-t^2\right)^2\right)^2}d\rho,
\ee 
and after integration we get the correction to the entanglement entropy
\be 
\Delta S \sim  -\frac{Q^2 }{8 \alpha}\left(\frac{\left(3 \alpha ^4+2 \alpha ^2 \left(3
   t^2+\ell ^2\right)+3 \left(\ell ^2-t^2\right)^2\right)
   }{\ell 
   \sqrt{\left(\alpha ^2+\ell ^2\right)^2+2 t^2 (\alpha^2 -\ell^2 )   +t^4}}\text{arccsch}\left(\frac{\left| \alpha ^2+t^2-\ell^2\right| }{2 \alpha  \ell }\right)-6 \alpha \right).
\ee 
In Fig.\ref{fig:EEQ3d-quench}\footnote{Here  $Q^2<0$ corresponding to the imaginary chemical potential is presented for completeness.} we present the evolution of entanglement entropy corresponding to the quench in $d=3$ CFT by local charged operator. 
Qualitatively  the entanglement evolution resembles the picture of $d=2$ case. The dip is more narrow and sharply peaked and, one should note that the behaviour is more monotonous without additional maximum and minimum. This behaviour is universal for $d>3$.

\section{Quench by the charged operator at finite temperature in 2d CFT}\label{sec:T}
\begin{figure}[t!]
\centering
\includegraphics[width=9.5cm]{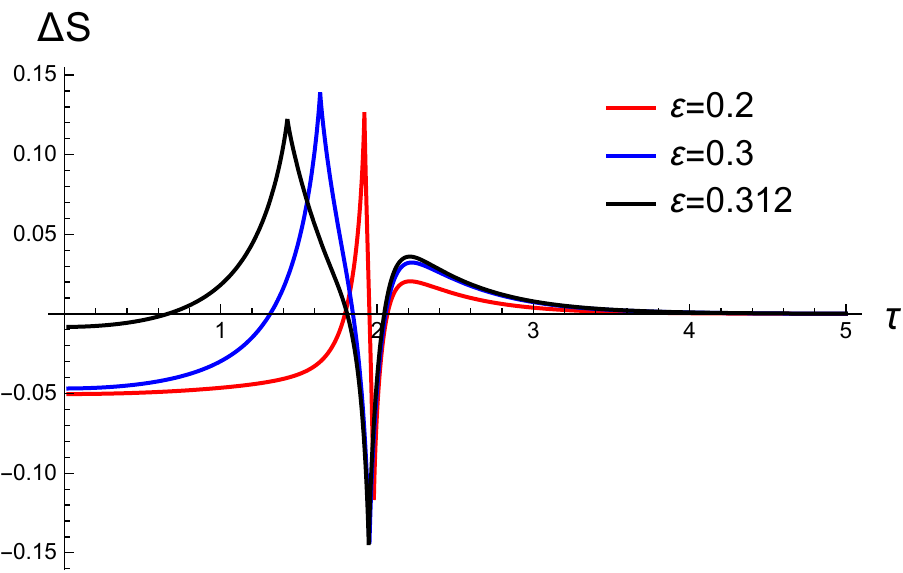}
 \caption{The entanglement entropy evolution for interval of length $\ell$  after  local charged operator quench at finite temperature for different values of $\varepsilon$. Here $M=1$, $\mu=0.1$, $Q=0.8$, the interval is centered around quench point $x=0$ and $\ell=4$. }
 \label{fig:EEQfint}
\end{figure}

The operator local quench in 2d CFT admits finite temperature extension in the straightforward manner  both on the gravity and CFT sides \cite{Caputa:2014eta}. The holographic finite temperature local quench construction is based on the mapping of the metric \eqref{cBTZ} to the BTZ frame by analogy with $T=0$ considered in the previous section. The explicit form of the map between Poincare  and BTZ frames is\footnote{Here $\varepsilon$ corresponds to $\alpha$ from the previous sections.}
\bea\label{coordtr}
&&\phi =\arctan \left(\frac{\varepsilon \sqrt{M} \sinh \left(\sqrt{M}
   x\right)}{\sqrt{1-M z^2} \cosh \left(\sqrt{M}
   t\right)-\sqrt{1-\varepsilon^2 M} \cosh \left(\sqrt{M}
   x\right)}\right),\\ \nonumber
&&\tau = -\arctan\left(\frac{\varepsilon \sqrt{M} \sqrt{1-M z^2} \sinh
   \left(\sqrt{M} t\right)}{\sqrt{1-\varepsilon^2 M} \sqrt{1-M z^2}
   \cosh \left(\sqrt{M} t\right)-\cosh \left(\sqrt{M}
   x\right)}\right),\\ \nonumber
&&R=\frac{L }{2 \varepsilon M z}\sqrt{{\cal A}_1+{\cal A}_2},\\ \nonumber
  &&{\cal A}_1=-8 \sqrt{1-\varepsilon^2 M} \sqrt{1-M z^2}
   \cosh \left(\sqrt{M} t\right) \cosh \left(\sqrt{M} x\right)-4
   \varepsilon^2 M+3-M z^2,\\ \nonumber
   &&{\cal A}_2=\left(2-2 M z^2\right) \cosh ^2\left(\sqrt{M}
   t\right)+\left(1-M z^2\right) \cosh \left(2 \sqrt{M} t\right)+2
   \cosh \left(2 \sqrt{M} x\right),
\eea
and for $\cM=0$ and $Q=0$ this map brings the metric \eqref{cBTZ} to the form of static one-sided BTZ black hole
  \bea\label{pbtz}
&&ds^2=\frac{L^2}{z^2}(-f(z)dt^2+\frac{dz^2}{f(z)}+dx^2),\\
&&f(z)=1-Mz^2,\,\,\,\, T=\frac{1}{2 \pi z_h},\,\,\,\,M=1/z_h^2.
 \eea
 For $Q\ne 0$ and $\cM\ne 0$, we get the dynamical metric corresponding to the BTZ black hole perturbed by the charged point-like object falling on the  horizon. The metric corresponding to this geometry  is  of a complicated form, so we will not write down it here explicitly as well. The extension of the results from the previous sections is straightforward. We use the approximation \eqref{apprS} where the geodesic corresponding to the unperturbed metric has the form
 \be 
 z_{BTZ}(x,\ell_1,\ell_2)=\frac{2e^{\sqrt{M} \left(\ell_1+\ell_2\right)/2} }{\sqrt{M}}\frac{ \sqrt{\sinh
   \left(\sqrt{M} \left(x-\ell_1\right)\right) \sinh \left(\sqrt{M} \left(\ell_2-x\right)\right)}}{e^{\sqrt{M} \ell_1}+e^{\sqrt{M} \ell_2}}.
 \ee 
We present the effect of the finite temperature on evolution of the entanglement entropy perturbation in  Fig.\ref{fig:EEQfint}. We see that the evolution  is more complicated in comparison with the zero temperature case. One can observe the presence of two asymmetric peaks surrounding the entanglement dip at $t\approx \ell/2$. The earlier peak  is sharp and the later one is more smooth. 

\section{Concluding remarks and summary}\label{sec:rem}
In this paper, we studied the dynamics of the holographic system after  the local charged operator quench. As a holographic dual we take the ``falling black hole'' construction of \cite{Horowitz:1999gf} applied to charged AdS black hole and extending the analysis of \cite{Nozaki:2013wia}. In $d=2$ case, this corresponds to the charged particle falling in the bulk of $AdS_3$ and deforming it. We focused on the study of the chemical potential, currents, and the entanglement entropy evolution. Let us briefly summarize our results.
\begin{itemize}
    \item We derived the explicit expressions describing charge-density waves after the quench. The dynamical chemical potential and charge evolution in a two-dimensional system and in higher dimensions are slightly different. In $d=2$ case, we see propagating charge lumps that change their sing after some time. For higher dimensions, this propagation is monotonous and seems to be universal for all dimensions $d>2$.
    \item The entanglement entropy dynamics for all dimensions have the universal feature. At some characteristic timescale, the system shows a sharp dip in the entanglement entropy. For $d=2$  the entanglement shows additional local mild maxima. 
    \item Finite temperature modifies the entanglement evolution and adds a sharp ramp before the dip.
\end{itemize}
It would be interesting to obtain the generalization of this setup in the case of Lifshitz-like space. Another interesting direction to explore is to consider charged AdS black holes perturbed by the charged particle. For a different  background chemical potentials particle will fall on the black horizon or oscillate between boundary and horizon. This bulk picture should describe some transition from the dissipative to oscillating dynamics in the dual quantum system like in \cite{Krikun:2019wyi}. 

\section*{Acknowledgements}
I would like to thank Andrey  Bagrov, Pawel Caputa, Mikhail Katsnelson, Alexander Krikun,  and Tadashi Takayanagi for discussions. Also I would like to thank Yulia Ageeva and Mikhail Khramtsov for comments on earlier versions of the manuscript.
This research is supported by the Foundation for the Advancement of Theoretical Physics and Mathematics
``BASIS'' (Project No. 18-1-1-80-4).
\newpage

\appendix

\section{The explicit form of the metric dual to neutral local quench}\label{app:metric}
 The explicit form of the metric (of reasonable size) can be obtained only for $M=0$.
The global $AdS_3$ metric deformed by a static neutral point particle of mass $m$ is given by
\be \label{appmetr1}
ds^2=-d\tau^2
   \left(L^2-{\cal  M}+R^2\right)+R^2 d\phi^2+\frac{L^2 dR^2 }{L^2-{\cal  M}+R^2},
\ee
where ${\cal  M}=8mGL^2$. For simplicity we take $L=1$.

Then, we obtain the holographic dual of the local quench by applying to \eqref{appmetr1} the following coordinate transformation
\bea\label{appmap}
&&\phi =\arctan\left(\frac{2 \alpha  x}{\alpha ^2+t^2-x^2-z^2}\right),\\
&&\tau =\arctan\left(\frac{2 \alpha  t}{\alpha ^2-t^2+x^2+z^2}\right),\\
&&R=\frac{\sqrt{\alpha ^4+2 \alpha ^2, \left(t^2+x^2-z^2\right)+\left(-t^2+x^2+z^2\right)^2}}{2 \alpha  z}.
\eea
After some algebra one can get this metric in the form
\bea \label{eq:long_metric}
&&ds^2=\frac{1}{z^2}\frac{\left(\alpha
   ^2 dx-2  t x dt+dx \left(u-z^2\right)+2 x z dz
   \right)^2}{\alpha ^4+2 \alpha ^2
   \left(u-z^2\right)+\left(z^2-v\right)^2}-\\\nn
&&-\frac{1}{z^2}\frac{\left(\alpha ^4+2
   \alpha ^2 \left(u+z^2(1-2 \mu) \right)+\left(z^2-v\right)^2\right)
   \left(\alpha ^2 dt+ \left(u+z^2\right)dt-2 t (xdx
   +zdz )\right)^2}{\left(\alpha ^4+2 \alpha ^2
   \left(u+z^2\right)+\left(z^2-v\right)^2\right)^2}\\\nn
&&\frac{1}{z^2}\frac{\left(\alpha ^4 dz+2 \alpha ^2 (udz -z(t dt +x dx ))+\left(v-z^2\right) \left(-2 t z dt +2 x z dx +\left(v+z^2\right)dz
   \right)\right)^2}{\left(\alpha ^4+2 \alpha ^2
   \left(u-z^2\right)+\left(z^2-v\right)^2\right) \left(\alpha ^4+2
   \alpha ^2 \left(-2 \mu
   z^2+u+z^2\right)+\left(z^2-v\right)^2\right)},
\eea
where we introduced $u=t^2-x^2$ and $v=t^2+x^2$. 

\section{Static charged backgrounds and the geodesic length}\label{sec:statApp}
\begin{figure}[t!]
\centering
\includegraphics[width=8.5cm]{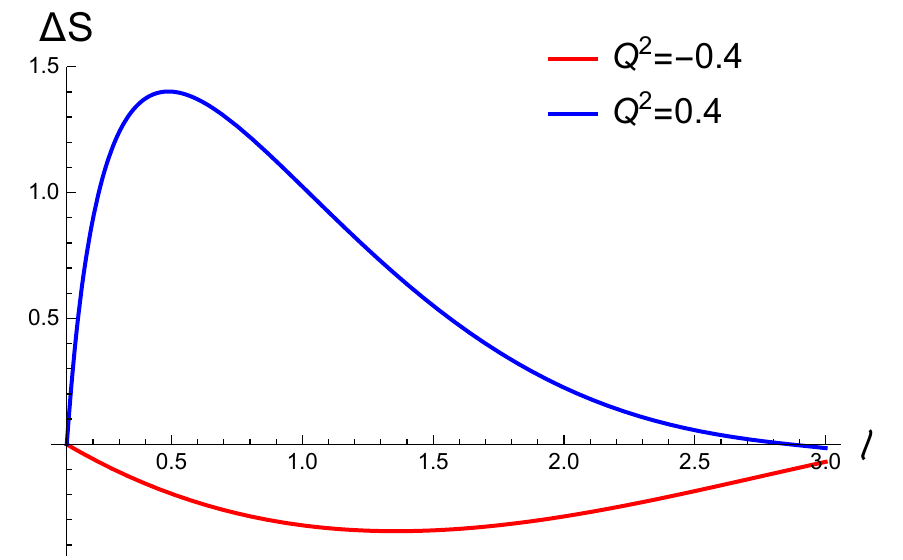}
 \caption{The entanglement perturbation $\Delta S=S(\ell,\cM,Q)-S(\ell,\cM,0)$ for $\cM=0.2$ and $Q^2=\pm 0.4$ for charged BTZ black hole.}
 \label{fig:EEQcbtz}
\end{figure}

In this Appendix we consider the effect of the charge on the geodesic length connecting two boundary points of metrics corresponding to \eqref{cBTZ0} and \eqref{blc}.
The standard Hubeny-Ryu-Rangamani-Takaynagi (HRRT) prescription states, that the entanglement entropy of the subregion $\phi \in (-\ell/2,\ell/2)$ is given by the minimal surface (in $d=2$ it is the geodesic)
\be 
S_{EE}=\frac{\cA}{4G}.
\ee 
We parametrize the geodesic by $R=R(\phi)$ with $R(\pm\ell/2)=\infty$.
The geodesic has turning point such that $R(0)=R_*$ and $R'(0)=0$. The parametric expression for the geodesic length $\cA$ has the form
\be \label{Af}
\cA(R_*)=2\int_{R_*}^{R_{reg}}\frac{R}{\sqrt{R^2-R_*^2} \sqrt{f(R)}},
\ee 
and the interval size  $\ell$ is
\be \label{Lf}
\ell(R_*)=2\int_{R_*}^{\infty}\frac{R_*}{R\sqrt{\left(R^2-R_*^2\right)f(R)}}.
   \ee 
Using these expressions one can obtain the entanglement entropy  for fixed $\ell$, $\cM$ and $Q$
\be
S=S(\ell,\cM,Q).
\ee 
We present the dependence of $\Delta S=S(\ell,\cM,Q)-S(\ell,\cM,0)$ in Fig.\ref{fig:EEQcbtz}.  We see, that the excitation of the entanglement entropy is non-monotonous for certain sign of $Q^2$.

$\,$

It is worth to comment the last choice corresponding to the imaginary  gauge fields and consequently the imaginary chemical potential values. Let us mention where the imaginary chemical potential takes place in physics. For example, the imaginary chemical potential have been studied in the context of charged Renyi entropies in \cite{Belin:2013uta}. The Gross-Neveu model and phase diagrams with imaginary chemical potential has been studied in \cite{Filothodoros:2016txa,Christiansen:1999uv,Karbstein:2006er}.

\begin{figure}[t!]
\centering
\includegraphics[width=8.5cm]{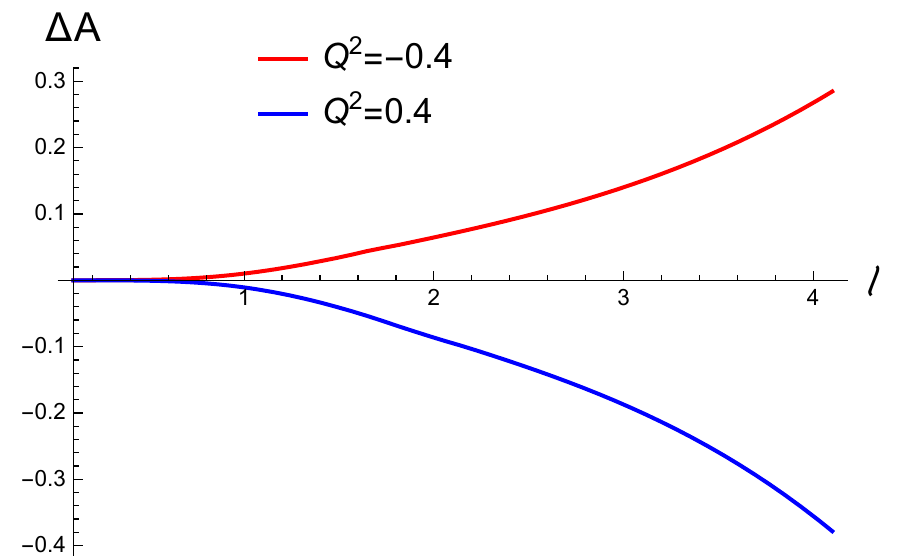}
 \caption{The length of geodesic spanned on the boundary interval of the length $\ell$   for $\cM=0.2$, $Q^2=\pm 0.4$  in higher-dimensional black holes where $f(R)$ is given by \eqref{blc}(with the same ). }
 \label{fig:EEQ3d}
\end{figure}

The calculation of the entanglement entropy  gives the real-valued answer also for $Q^2<0$ for background by \eqref{cBTZ0}. 
We compare the behaviour of the geodesic length spanned on the interval of length $\ell$ in $d=2$ and in the geometry defined by the function  \eqref{blc}. This quantity does not give the entanglement in higher-dimensional case and is relevant to the equal-time Green function. However, it can give some intuition to correlations in the dual defined by \eqref{blc}. We present the geodesic length difference between charged and chargeless case in Fig.\ref{fig:EEQ3d}. We choose $d=3$ such that $d\Omega=d\theta^2+\sin^2 \theta d\phi^2$ and fix  $\phi=const$. 
In contrast with $d=2$, it exhibits the monotonous dependence.

\section{The charged fields after the mapping}\label{sec:gauge}
The gauge field corresponding to the dynamical background describing charged local quench can be obtained in the straightforward manner  applying the mapping \eqref{mapRz} to the static gauge field \eqref{btzgauge} 
\bea \label{dyngauge}
\nn
&&A=A_{t}dt+A_{x}dx+A_{z}dz,\,\,\,\,\,\,\,\,\,\,Q_{eff}=\Big(\tilde Q_{eff}(t,x)+\log z \Big),\\ \nn
   &&Q_{eff}(t,x)=Q\log \left(2 \alpha R_h \sqrt{\alpha ^4+2 \alpha ^2
   \left(t^2+x^2-z^2\right)+\left(-t^2+x^2+z^2\right)^2} \right),\\
&&A_t=\frac{2 \alpha   \left(\alpha ^2+t^2+x^2+z^2\right)
   }{\alpha ^4+2 \alpha
   ^2
   \left(t^2+x^2+z^2\right)+\left(-t^2+x^2+z^2\right)^2} \cdot Q_{eff}(t,x),\\
&&A_x=-\frac{4 \alpha   t x }{\alpha ^4+2 \alpha
   ^2
   \left(t^2+x^2+z^2\right)+\left(-t^2+x^2+z^2\right)^2}\cdot Q_{eff}(t,x),\\
 &&A_z=-\frac{4 \alpha   t z }{\alpha ^4+2 \alpha
   ^2
   \left(t^2+x^2+z^2\right)+\left(-t^2+x^2+z^2\right)^2}\cdot Q_{eff}(t,x).
\eea 
\section{The evolution of the metric on the constant time slice}

\begin{figure}[h!]
\centering 
\includegraphics[width=8.5cm]{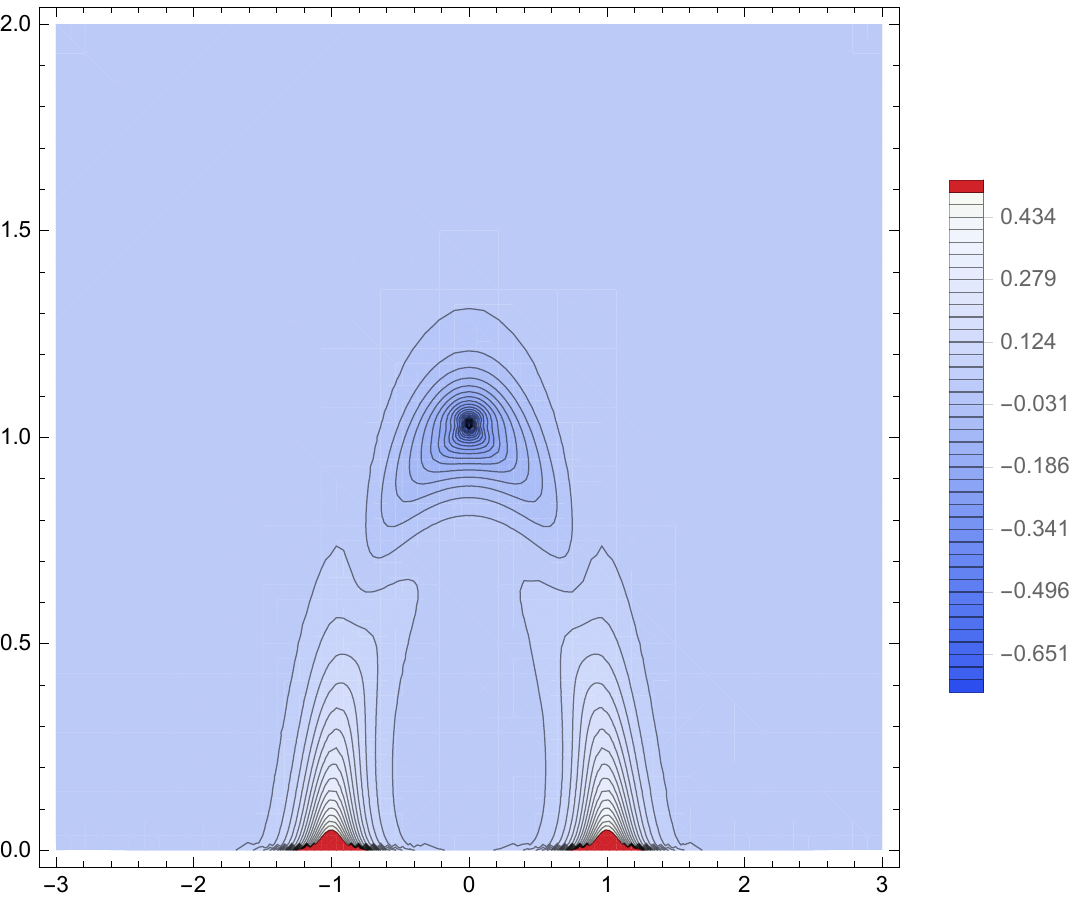}
 \caption{The determinant of the metric induced on the constant time slice. We take the difference between the values corresponding to the $Q=0.1$ and $Q=0$ of this determinant. For $M=0.1$ and fixed time moment $t=1$. We see that  there is a negative contribution (colored by blue) around the particle position, while in the UV region there is a positive contribution(red).}
 \label{fig:dens}
\end{figure}

\end{document}